\documentstyle[12pt,aaspp4]{article}

\def\gtaprx{\mathrel{\vcenter{\offinterlineskip \hbox{$>$}
    \kern 0.3ex \hbox{$\sim$}}}}
\def\ltaprx{\mathrel{\vcenter{\offinterlineskip \hbox{$<$}
    \kern 0.3ex \hbox{$\sim$}}}}

\renewcommand{\sun}{\odot}
\newcommand{\Msun}{\mbox{$M_\sun$}}

\begin{document}

\title{THE LUMINOSITY FUNCTION AND INITIAL MASS FUNCTION IN THE GALACTIC BULGE
\footnote{Based 
on observations with the NASA/ESA {\it Hubble Space Telescope}, obtained at 
the Space Telescope Science Institute, operated by AURA Inc under contract 
to NASA}}

\author{
Jon~A.~Holtzman\altaffilmark{2},
Alan~M.~Watson\altaffilmark{3},
William~A.Baum\altaffilmark{4},
Carl~J.~Grillmair\altaffilmark{5},
Edward~J.Groth\altaffilmark{6},
Robert~M.Light\altaffilmark{7},
Roger Lynds\altaffilmark{8},
Earl~J.O'Neil, Jr.\altaffilmark{8}
}

\altaffiltext{2}{Department of Astronomy, New Mexico State University, Dept 4500 Box 30001, Las Cruces, NM 88003, holtz@nmsu.edu, awatson@nmsu.edu}
\altaffiltext{3}{Instituto de Astronom\'\i a UNAM, J. J. Tablada 1006, Col. Lomas de Santa Maria, 58090 Morelia, Michoac\'an, Mexico, alan@astrosmo.unam.mx}
\altaffiltext{4}{Astronomy Department, University of Washington, Seattle WA 98195, baum@astro.washington.edu}
\altaffiltext{5}{Jet Propulsion Laboratory, 4800 Oak Grove Drive, Pasadena, CA 91109, carl@wfpc2-mail}
\altaffiltext{6}{Department of Physics, Princeton University, Princeton, NJ 08544, groth@pupgg.princeton.edu}
\altaffiltext{7}{Caltech/IPAC, MS 100-22, Pasadena, CA 91125, light@ipac.caltech.edu}
\altaffiltext{8}{Kitt Peak National Observatory, Box 26732, Tucson AZ 85726,
lynds@noao.edu, oneil@noao.edu}

\begin{abstract}

We present deep photometry obtained with the Hubble Space Telescope
(HST) in a field in Baade's Window in the Galactic bulge. We derive a
luminosity function down to $I\sim 24.3$, or $V\sim 27.5$,
corresponding to $M\sim 0.3$\Msun. The luminosity function from the
turnoff down to this level appears remarkably similar to that observed
in the solar neighborhood. We derive a mass function using both an
empirical local mass-luminosity relation and a mass-luminosity relation
from recent stellar model calculations, allowing for the presence of
binaries and photometric errors. The mass function has a power law form
with $dN/dM\propto M^{-2.2}$ for $M\gtaprx 0.7$\Msun. However, we find
strong evidence for a break in the mass function slope around 0.5-0.7
\Msun, with a significantly shallower slope at lower masses. The value
of the slope for the low masses depends on the assumed binary fraction
and the accuracy of our completeness correction.
This mass function should directly reflect the {\it initial} mass
function.

\end{abstract}

\section{Introduction}

The luminosity function of low mass stars is particularly interesting
because it can be used to infer the initial mass function (IMF)
independently of the star formation history, since low mass stars
evolve relatively little over the entire lifetime of the Universe.
Understanding whether or not there are variations of the IMF with
galaxy type or with metallicity is essential to our understanding of
star formation and to the modelling of galaxy evolution. To date, most
observations of IMFs of low mass stars have been made, by necessity, in
nearby systems.

In the solar neighborhood, estimates of the low-mass IMF have been made
by Salpeter (1955), Miller \& Scalo (1979), and Kroupa, Tout, \&
Gilmore (1993) (among others). The latter find that a segmented power
law, $dN/dM\propto M^{\alpha}$ best represents the local IMF, with
$\alpha=-2.7$ for $M>1$\Msun, $\alpha=-2.2$ for $0.5<M<1$\Msun, and
$-1.85<\alpha<-0.7$ for $M<0.5$\Msun.  The mass function of the least
massive stars is of particular interest in connection with the
frequency of brown dwarfs, the local mass density in the Galactic disk,
and the observed microlensing rates towards the LMC and the Galactic
bulge. Other recent studies of the local neighborhood also find a
flattening of the IMF slope at low masses; Gould, Bahcall, \& Flynn (1997) find
$\alpha = -0.56$ for $M<0.5$\Msun\  for HST observations of local M
dwarfs. The Gould et al. slope does not include a correction for
the presence of binaries; allowing for these brings the inferred
slope into the range suggested by Kroupa et al. (1993), who did make
a correction for binaries.

Measurements of the luminosity and mass functions in the Galactic halo
have been recently reviewed by Mould (1996); conflicting results have
been reported.  Dahn et al. (1995) find a turnover in the halo
luminosity function, while Richer \& Fahlman (1992) find evidence for a
rapidly increasing number of stars as one goes to lower masses. In support
of the Dahn et al. result, Gould, Flynn, \& Bahcall (1997)
derive a mass function with a shallow slope of $\alpha \sim -0.75$
from analysis of 166 spheroid subdwarfs observed with HST.

Other estimates of IMFs have been made in stellar clusters and
associations, both in the Galaxy and in some nearby stellar
systems.  Most of these measurements suggest IMFs similar to those
observed in the solar neighborhood, although there are some exceptions;
Hunter et al. (1997) present a recent summary. However, measurements of
stars with $M<0.5$\Msun\ have been difficult to make. DeMarchi \&
Paresce (1995a, 1995b, 1997) have measured the luminosity function down
to very low mass stars in several nearby globular clusters using HST,
and they find an increasing number of stars with decreasing mass down
to $\sim$ 0.2 \Msun, but then a flattening of the luminosity function
towards lower masses. However, this is in conflict with previous
ground-based measurements in these clusters, which suggest steeply
rising mass functions down to the lowest mass stars observed (Richer et
al. 1991). In general, the connection between current cluster mass
functions and their initial mass function may be complicated because of
dynamical evolution within a cluster and the removal of stars by the
tidal field of the parent galaxy.

In nearby galaxies, the ability of HST to observe individual faint
stars has allowed estimates of the initial mass function down to
relatively low stellar masses ($M\sim 0.7$ \Msun). However, the
conversion between a luminosity function and a mass function is not
totally independent of the star formation history for such stars because
these stars evolve in luminosity in less than a Hubble time.  In
an outer field in the Large Magellanic Cloud (LMC), Holtzman et al.
(1997) find that a solar neighborhood initial mass function provides an
adequate match to the data, but only if there is a significant
component of older stars in the LMC. If the stars in the LMC field are
predominantly young, then a steeper IMF, with $\alpha \sim -2.75$, is
required. In the nearby dwarf spheroidal Draco, the inferred IMF is
similarly linked to the age of the system (Grillmair et al. 1998); for
an age of 12 Gyr, the inferred IMF slope is comparable to that of the
solar neighborhood.

Consequently, a picture is emerging which suggests, remarkably, that the
initial mass function does not appear to vary significantly from one
environment to another. However, much of the interpretation is still
complicated by lack of knowledge about star formation histories, which
affect inferences about the initial mass function for all except very
low mass stars, and by the possible effects of dynamical evolution in
star clusters. Additionally, initial mass functions have not been
measured for all types of stellar systems. In particular, no metal-rich
systems have been studied, nor have any massive spheroidal systems.

A determination of the initial mass function of the Galactic bulge for
comparison with that of the disk is important because of the
possibility of differing modes of star formation in spheroidal and disk
systems. Furthermore, a measurement of the mass function of stars in
the bulge is essential to interpretations of microlensing events
observed in the direction of the Galactic center (e.g., Alcock et al.
1997).

We have observed a field in Baade's Window with the Wide Field
Planetary Camera 2 on the Hubble Space Telescope in order to observe
faint, low mass, stars. Our observations probe to stars with $M\sim
0.25$ \Msun.  In this paper we concentrate on the luminosity function
of the faint stars, and its implications for the mass function in the
bulge. A subsequent paper  will discuss the interpretation of the
color-magnitude diagram in greater detail, concentrating on what it can
tell us about the star formation history in the Galactic bulge.

\section{Observations}

Observations of Baade's Window were obtained on 12 August 1994 with the
Wide Field Planetary Camera 2 of the Hubble Space Telescope.
Observations were made through the F555W and F814W filters (wide $V$ and
I) with a total of 2420s through each filter. Observations through each
filter were split into 5 exposures with exposure times of 20, 200, 200,
1000, and 1000 seconds.  To maximize the dynamic range, the short
exposures were made with a gain of $\sim$ 14 electrons/DN, and the long
exposures were made with gain $\sim$ 7 electrons/DN.

The data were processed using the standard reduction techniques
discussed by Holtzman et al. (1995a, H95A). This processing included a
very small correction for analog-to-digital errors, overscan and bias
subtraction, dark subtraction, a tiny shutter shading correction, and
flat fielding.

\subsection{Photometry}

Figure 1 shows the combined set of F555W exposures.  The field is
crowded and profile-fitting photometry is required to get accurate
results for the faint stars.

To perform the profile-fitting photometry, the five frames in each
color were first combined to reject cosmic rays based on the known
noise properties of the WFPC2 detectors. Since the F814W frames were
deeper (because we are probing to low mass, very red stars), star
detection was performed on the F814W frame alone.  Given the input star
list, profile fitting was performed simultaneously on the set of ten
individual frames, solving for a brightness for each star in each
color, a position for each star, a separate background value for each
group of stars in each frame, and frame-to-frame pointing shifts.
Simultaneous fitting imposes the requirement that all frames have the
same star list with the same relative positions (after allowing for the
variation in scale as a function of wavelength as discussed in H95A).
Small pointing differences between the frames provide slightly
different pixel samplings of the PSF, providing additional information
for fitting the undersampled PSF.  Model PSFs that vary across the
field of view were used; separate models were derived for each of the
individual frames allowing for small focus shifts between frames.  A
brief description of the model PSFs and their advantages and
disadvantages is presented in Holtzman et al. (1997).  Cosmic rays in
each of the individual frames were flagged by the procedure which
combined the stack of frames for star-finding, and contaminated pixels
were ignored in the profile-fitting procedure.

Because of the large range in luminosities of stars observed in the
bulge, the wings of the bright stars cause significant problems for
automatic star finding algorithms.  To minimize the problem, the
profile fitting was iterated three times. In the first pass, only the
brightest stars were fit. This allowed identification and subtraction
of these stars including the extensive stellar wings and diffraction
spikes.  In the second pass, the star finding algorithm was used on the
subtracted frames, with a low threshold to detect faint stars.  A
higher detection threshold was used around bright stars in the
subtracted frame to avoid spurious detections from imperfect PSF
subtraction. A fit was then performed on the original frames including
stars found on both first and second passes, and these stars were
subtracted. In the third pass, a few additional close neighbors of
stars were detected from these subtracted frames. These were added to
the list of stars, and a final stellar photometry run was made. During
each of the profile-fitting stages, the software attempted to remove
spurious detections by deleting stars that were not well fit by the
stellar PSF.  The final photometry list was filtered once again using a
goodness-of-fit index in an attempt to remove spurious detections which
remained.

The resulting magnitudes were placed on the synthetic WFPC2 photometric
system defined by Holtzman et al. (1995b, hereafter H95B).  The profile results
were converted to instrumental aperture magnitudes with a 0.5 arcsec
radius aperture using aperture photometry of reasonably bright stars
after subtraction of their neighbors based on the profile fitting
results. The aperture corrections were determined by inspecting the
difference between the 0.5 arcsec aperture and profile-fitting results;
a separate correction was determined for each of the four chips,
although they all agreed to within a few percent.  We judge the
accuracy of the aperture corrections to be a few percent in the worst
case. Because these were fairly long, crowded exposures, we made no
correction for possible errors from charge transfer efficiency (CTE)
effects, as discussed in H95B; if CTE
problems were present they would only change the derived magnitudes by
a few percent and our conclusions would be unaffected. No correction
was made for a possible systematic effect which may give differences in
photometric zeropoint between long and short exposures (see Note Added
in Proof, H95B); applying such a correction would make all our
magnitudes about 0.05 mag fainter, which would also have a minimal
impact on our conclusions.

To compare with local luminosity functions which have been derived in
the $V$ and $I$ bandpasses, we also transformed our WFPC2 magnitudes to the
Johnson/Cousins $VI$ system using the synthetic transformations presented
in H95B. These transformations were derived from a stellar library
which included stars as red as those observed here.  The use of these
transformations introduces some potential systematic errors because of
the unknown metallicity dependence of the transformations, but such
errors are likely to be small. In any case, most of the work presented
below is performed in the native WFPC2 system.

The calibrated color-magnitude diagrams (with both F555W and F814W on
the ordinate) are presented in Figure 2.  A well defined main sequence
can be seen down to $\mathrm{F555W}\sim V>27$.

\subsection{Completeness and Error Estimation}

To accurately interpret the luminosity function, we need to understand
the detection efficiency and measurement errors as a function of
stellar brightness. To estimate these, we performed a series of
artificial star experiments in which we added a grid of stars of equal
brightness onto each exposure in each of the four chips. Artificial
stars were given colors corresponding to the median color of observed
stars at a comparable magnitude, so fainter stars were made to be
redder.  The grid spacings were chosen to insure that the artificial
stars were isolated from each other and thus did not add significantly
to the crowding on the frame; 529 stars were placed on each of the WFs,
and 121 were placed on the PC.  Different pixel centerings were used
for each artificial star, and the pixel centering varied slightly from
frame-to-frame as in the real data.  Poisson statistics were used to
add errors to the artificial stars.  These frames were then run through
photometry routines identical to those discussed in Section 2.1.  This
was done 22 separate times with different brightnesses chosen for the
artificial stars each time.

For each of the artificial star runs, the final list from the
photometry procedure was compared with the input list of artificial
stars, and also with the final photometry list from the original
frames. An artificial star was considered to be found if there was a
detection within one pixel of the position where the star was placed
and if there was no corresponding detection on the original frame. If a
match was found with both the artificial star position and with an
object on the original frame, the artificial star was considered found
if the measured F555W magnitude was closer to the magnitude of the
artificial star than to the magnitude of the star on the original
frames. This properly accounts for incompleteness arising from crowding
as well as from incompleteness from inability to detect stars in the
noise of the background.

The artificial stars also provided an estimate of the photometric
errors, at least for the fainter stars.  A limitation is that the
artificial stars are created and measured with the same PSF, so there
are no errors resulting from inaccuracies in the PSF models. Such
errors dominate for brighter stars, so the artificial stars cannot be
used to judge the photometric errors for these stars.  Errors in the
fainter stars are dominated by photon statistics and include both
random and systematic errors.  The former comes from Poisson statistics
and readout noise, but systematic errors also occur at the faintest
levels because objects with positive noise fluctuations are detected
preferentially over those with negative fluctuations. Systematic errors
can also arise from crowding.

Some of the measured completeness and error distributions for the F814W
magnitudes are shown in Figure 3. Each panel shows a histogram of
observed errors for artificial stars of a different brightness.  The
text in each panel identifies the artificial star brightness (F555W and
F814W) as well as the completeness fraction (fraction of artificial
stars detected and measured).  As expected, the random error increases
for fainter objects. In addition, for the faintest objects, it is clear
that the error distribution is asymmetric for the F814W magnitudes,
with more stars being detected too bright than too faint. This is
expected, since the faintest objects may only be detected if they have
a positive noise fluctuation. Crowding may also contribute to this
result.

The accuracy of the completeness tests is different for the two
different filters, because star detection is performed only on the
F814W frames.  The probability that an artificial star will be detected
depends on its F814W magnitude. Consequently, completeness as a
function of input F814W magnitude is accurately measured, but
completeness as a function of input F555W magnitude is accurate only to
the extent to which the artificial stars have the {\it same color} as
the true stars.  The artificial star colors were chosen based on median
colors of the observed real stars, but these are likely to be biased
for the faintest stars by incompleteness. Consequently, we believe that
the F814W corrected luminosity function is more accurate than the
corresponding function in F555W. In addition, random errors are smaller
in F814W for the faintest stars, so smearing of the luminosity function
from observational error is smaller in the F814W luminosity function.

We have attempted to assess possible errors in our completeness
corrections by repeating the test with simulated stars made using a PSF
that has a severely different focus from that inferred from the actual
frames. We then reduced these frames with our normal PSFs to simulate
the effect of using an erroneous PSF. Completeness results for the two
different PSFs are shown in Figure 4.  Although these differ
significantly for the faint objects, we note that in the repeat test we
used a PSF for the fake objects which was an extreme mismatch; the
subtractions from our incorrect PSFs were glaring and far worse than
any subtractions of comparably bright real stars.  Consequently, we
feel that the differences illustrated between these two completeness
curves represent the extreme of possible errors.

Spurious detections are more problematic than missed detections, since
it is more difficult to estimate their frequency.  We have attempted to
minimize the number of spurious detections by using a  relatively high
star finding threshold, and by using a conservative limit on
goodness-of-fit for accepting objects for which we perform photometry.
Visual inspection shows that we do not appear to have a large number of
spurious detections remaining after these techniques are applied; in
subsequent analysis, we make no effort to correct for the few which
have survived, since we cannot determine a reasonable estimate for the
number of spurious detections as a function of apparent magnitude.

\section{Interpretation}

\subsection{Distance and reddening}

We have adopted a distance of 8 kpc for the bulge (see, e.g., Carney et
al.  1995), with a corresponding distance modulus of 14.52. Maximum
errors in this are probably about 0.3 mag (corresponding to distances
between 7 and 9 kpc). Of course, the bulge population sampled is likely
to lie at a range of distances, causing the observed luminosity
function to be smeared.

The extinction in the direction of Baade's Window has been discussed by
Stanek (1996), Gould, Popowski, \& Terndrup (1998), and Alcock et al.
(1998), among others. Stanek (1996) presents a map of differential
reddening within Baade's Window based on brightnesses of red clump
stars. Gould et al. (1998) have computed a zero point for this map
based on observed $(V-K)$ colors as compared with $(V-K)_0$ predicted
from observed H$\beta$ indices. Alcock et al. (1998) independently compute a
zeropoint based on observations of RR Lyrae stars and derive an almost
identical zeropoint to that of Gould et al. Using this zeropoint and
the Stanek map, we infer an extinction of $A_V=1.28 \pm 0.08$ for our
field, which lies in one of the clearest regions of Baade's Window.
Using the calculations of Holtzman et al.  (1995b), we infer
extinctions in the WFPC2 filters system of $A(\mathrm{F555W})=1.26$ and
$A(\mathrm{F814W})=0.76$.


\subsection{The luminosity function}

Figure 5 shows the observed luminosity function in the $V$ and $I$ bands.
Both the uncorrected (open squares) and completeness-corrected (filled
squares) luminosity functions are shown.  The completeness correction
here uses the completeness fraction as measured from simulated stars of
the corresponding magnitude.  The application of the completeness
correction in this way is only approximate because it assumes that
stars are measured without observational error; in reality, stars
observed at a given magnitude actually have a range of true magnitudes,
and, correspondingly, different detection probabilities.  However,
random errors are not likely to have much effect because the luminosity
function is relatively flat, and, as shown above, systematic errors are
not important until $V\sim 28$, or $M_V\sim 12$.  Similarly, spread in
the distance of the stars is not likely to significantly affect the
relatively flat luminosity function.

We show the corrected luminosity function for $M_V < 12.25$ and
$M_I<9$; as discussed above, the completeness correction for $I$ is
probably more reliable than that for $V$ because the latter depends on
the accuracy of the simulated star colors. Errors in the completeness
correction are a likely cause of the apparent turnover in the $V$ band
luminosity function at the faintest magnitudes. For the $I$ band, we expect
that the error in completeness gives an uncertainty of
$\ltaprx$ 50\% in the counts at the very faintest magnitude shown.

For comparison, we also show the solar neighborhood luminosity function
from Wielen et al. (1983) (triangles), as well as a recent
determination of the local luminosity function for M dwarfs as derived
by HST imaging (asterisks, Gould et al. 1997).  The luminosity
functions have been normalized to agree at $M_V=9$ and $M_I=7.25$. For
the $I$ band, these luminosity functions have been transformed from the V
band using the relation between $V$ and $V-I$ presented by Kroupa \& Tout
(1997) based on the data of Monet et al (1992).

It is immediately apparent that the corrected bulge luminosity function
is in close agreement with the solar neighborhood luminosity function
over the range $7<M_V<11$ and $6<M_I<9$. Brighter than this, the bulge
luminosity function drops off more steeply than the local function, as
expected for an older population. The one discrepant point from the
Gould et al. luminosity function ($M_V\sim 8.3$) has a large associated
error because few stars this bright are counted in the HST fields.

However, the match of the luminosity function with that of the solar
neighborhood does not necessarily imply a correspondence in the mass
functions because of possible differences in the number of binaries in
the samples and because of observational errors. To consider these effects,
we turn to a discussion of the inferred mass function.

\subsection{The mass-luminosity relation}

The IMF is constrained using the lower main sequence because the
effects of stellar evolution are minimal for low mass stars over the
age of the universe. However, the derived mass function depends on an
accurate knowledge of the mass-luminosity relation, and calculations
indicate that the mass-luminosity relation depends on metallicity
(Kroupa \& Tout 1997). Theoretical mass-luminosity relations are
difficult to calculate for low mass stars because of complications from
the equation of state, opacities, and convection, leading to
uncertainties in the mass-luminosity relation as derived from models.
However, recent progress has been made by Baraffe et al. (1997).  These
models incorporate the most up-to-date physics available and are
computed self-consistently with the stellar atmospheres of Allard et
al.  (1997). So far, we have obtained these models only for stars up to
0.7 \Msun; for more massive stars (which do not enter strongly into the
discussion in this paper), we have used models from the Padua group
(Bertelli et al. 1994; Bressan et al. 1993; Fagotto et al. 1994a,b).  A
good summary of the current understanding of mass-luminosity relations
is presented by Kroupa \& Tout (1997). We note that it is clear that
the models are still not perfectly accurate, because the model 
color-magnitude relation falls blueward for the data for the faintest
stars.

Since uncertainties about the quality of the theoretical
mass-luminosity function remain, we also consider the use of an
empirical mass-luminosity relation. This is available only for the
solar neighborhood, and, consequently, only for stars of near solar
composition. However, the median metallicity observed in the bulge may
actually be quite similar to that of the solar neighborhood (McWilliam
\& Rich 1994), although the bulge metallicity distribution has a tail
which extends to lower metallicities.  Consequently, it is plausible
that an empirical mass-luminosity relation derived from solar
neighborhood stars will provide a reasonable match for the bulge. Such
empirical relations have been presented by Henry \& McCarthy (1993) and
Kroupa et al. (1993), and the two show good agreement.  However, the
Henry and McCarthy relation is presented as a series of quadratic fits
in different mass ranges. As a result, the derivative of their
function, which enters into the derivation of a mass function from a
luminosity function, is not continuous between the different regions
which they fit, leading to problems with its use.  Consequently, we
adopt the Kroupa et al. function as our empirical function.  This
function has been tabulated for both the $V$ and $I$ passbands (among
others) for stars with $M\leq 0.65$ by Kroupa \& Tout (1997).  For
larger masses, the relation is given by Kroupa et al.  (1993), but only
for the $V$ bandpass. 

To get the $I$ band mass-luminosity relation, we have transformed the V
band mass-luminosity relation to the $I$ band using a fit to
color-magnitude data for solar neighborhood stars which was kindly
provided by I.N. Reid; these data include ground-based measurements as
well as those from the Hipparcos satellite.  The applicability of this
relation to the bulge stars can be judged by the degree to which the
color-magnitude diagram of the bulge matches that of solar neighborhood
stars. Figure 6 shows the median locus of the bulge stars compared with
the solar neighborhood fit. One can see that these agree fairly well,
though not perfectly.  Minor differences may arise from different
metallicity distributions between the bulge and the solar neighborhood,
different fractions of binary systems in the samples, and
errors in our assumed distance and/or extinction. Figure 6 also shows
a solar metallicity model color-magnitude relation; this demonstrates
the problems the models have getting the correct colors for the fainter
stars.

Because neither the model nor the empirical mass-luminosity-color
relations match the observed properties of the bulge stars perfectly,
slightly different inferences are made about the mass function
depending on whether the $V$ or the $I$ band luminosity function is
considered.

\subsection{The IMF}

One can naively derive a MF from the luminosity function simply by
using the M-L relation to effect a change of variables.  However, this
method has no way of accounting for systematic or random errors in the
photometry which may be important for the fainter stars; it also cannot
account for the presence of binary stars or spread in the distance to
the stars. Of these different effects, the presence of binaries is the
most significant, especially for the low-mass stars considered here.
The effects of binaries have been previously discussed by Kroupa (1995)
and Kroupa, Tout, \& Gilmore (1993). Some of these effects are shown in
Figure 7, which plots expected luminosity functions for the {\it same}
initial mass function using our estimated completeness and several
different assumptions about the presence of binaries and systematic
errors. In this figure and hereafter, the binary fraction refers to the
number of \textit{systems} which are binaries.  Also, we make the
assumption that the masses of stars in binary systems are drawn
independently from the same mass function.  In Figure 7, the luminosity
functions have been normalized to match at the bright end to make the
differences in slope at the faint end most apparent. One can see that
the presence of binaries can have a severe effect on the observed
luminosity function. Depending on the mass function slope, this can
dominate over the relatively small effects that random and systematic
errors have on the luminosity functions, even for the faintest stars.

In the solar neighborhood, various studies suggest that the binary fraction
is in the vicinity of 0.5 (see discussion in Kroupa 1995 and Kroupa et al. 
1993). Of course, we have no idea whether the bulge binary fraction is
similar to that of the solar neighborhood, so we consider it to be a
free parameter.

To account for the presence of binaries and errors, a derivation of an
IMF involves simulating a luminosity function from some assumed mass
function, allowing for systematic errors, binaries, and distance
spread, and then checking for consistency with the observed luminosity
function. To do this, however, requires some parameterization of the
mass function in order to keep the number of possible models reasonably
small. Here, we initially transform our luminosity function into a mass
function ignoring binaries and errors in order to determine what might
provide a useful parameterization, and then simulate luminosity
functions with binaries and errors for a more sophisticated comparison
with model IMFs.

\subsubsection{No binaries or errors}

Figure 8 presents mass functions derived using both the F814W (top) and
the F555W (bottom) luminosity functions; results using the F555W are
more uncertain because the F555W data have larger photometric errors
and less accurate completeness estimates.  The inferred mass functions
are shown using the empirical mass-luminosity relation (squares), a
solar metallicity model mass-luminosity relation (triangles), and a
model mass-luminosity relation for a population with Z=0.006 ([Fe/H]
$\sim=-0.5$). For one of the relations (triangles), solid points show
completeness-corrected data and open points show raw data, to
illustrate the amplitude of the completeness corrections.

Independent of the choice of mass-luminosity relation, no single power
law mass function is able to fit the data; the derived mass-function
shows a break around 0.5-0.7 \Msun.  This conclusion depends on having
a reasonable estimate of the completeness, since the turnover occurs at
a level where our data are only $\sim$ 50\% complete.  However, our
completeness estimate would have to be off by a factor of two for stars
of 0.4 \Msun\  to be consistent with a single power-law mass function.
As discussed in \S 2.2, we do not believe this is likely.

For masses less than $\sim$ 0.7 \Msun, evolution is negligible, so this
result implies that the {\it initial} mass function cannot be fit with
a single power law.  A similar result is derived for the solar
neighborhood by Kroupa et al. (1993) and by Gould et al.  (1997). Both
of these studies find a mass function slope of $\alpha = -2.2$ for
stars with $M>0.5$ \Msun. For lower mass stars, Kroupa et al.  find a
slope of $-1.85<\alpha<-0.7$, and Gould et al. find $\alpha = -0.56$.
Lines in Figure 8 are shown which correspond to $\alpha = -2.2$ and
$\alpha = -0.56$. The data appear to be matched by a mass function
with a faint end slope of $\alpha > -1$.

The more massive stars are reasonably well matched by $\alpha=-2.2$ for
$M\gtaprx 0.7$ \Msun\ using the model mass-luminosity relation.  The
empirical mass-luminosity relation suggests a steeper slope for the
most massive stars, but since evolutionary effects are significant for
these stars, the empirical mass-luminosity relation is likely not
applicable since the mean age of bulge stars is larger than that of solar
neighborhood stars.

\subsubsection{The effect of binaries and errors}

As mentioned above, an accurate mass function cannot be derived by
simply converting luminosities to masses because of the presence of
binaries, systematic errors, and distance spread. Here we derive some
model luminosity functions assuming mass functions with power law
segments, motivated by the estimates provided by Figure 8.

The calculation of these models is complicated because of the
observational incompleteness. In principle, one should be able to take
the model magnitudes from a mass-luminosity relation, derive observed
magnitudes using a distance and extinction estimate, and use the
completeness estimate at that magnitude to predict an observed number of
stars. In practice, however, this leads to problems because any errors
in the mass-luminosity relation lead to large errors in the
completeness corrections. As mentioned above, it is clear that such
errors exist because neither the stellar model nor the empirical
mass-luminosity relations are able to match the observations
simultaneously in both bandpasses. The differences in the completeness
correction which one derives from using the two different bandpasses to
compute completeness can be severe. To avoid this problem, we compute
model luminosity functions without accounting for incompleteness in the
model, and compare these with the \textit{completeness-corrected}
data.  In this section, we only show comparisons with the F814W
luminosity function, which has the better determined completeness.

Figure 9 shows the observed luminosity function with calculated
luminosity functions assuming $\alpha = -2.2$ for $M>0.5$ \Msun\  and
$\alpha = -0.5,-0.9,-1.3,-1.7$ for $M<0.5$ \Msun.  These cover
the possible ranges of solar neighborhood mass functions inferred by
Kroupa et al. (1993). We also include a mass function with a constant
power law slope $\alpha=-2.0$ (steepest curve).  Results from both a
solar metallicity model and an empirical mass-luminosity relation are
shown (left and right), as well as results for three different binary
fractions, where the binary fraction gives the number of systems which
are binary. 

The top panels show results for no binaries, the middle panels for 50\%
binaries and the bottom panels for 90\% binaries, where binaries are
assumed to have uncorrelated masses.  As noted above, binaries have a
strong influence on the luminosity function of faint stars.  The models
with no binaries seriously overestimate the number of faint stars,
unless the faint-end slope flattens significantly at $M_{\mathrm{F814W}}\sim
6.5$, corresponding to $\sim$ 0.7 \Msun. With binaries, the models
provide a better match, although all models shown here are
statistically significantly different from the observed data. Including
binaries allows a steeper faint-end slope, but models with a constant
slope at $\alpha \ltaprx -2$ are inconsistent with the data.

The left panels use the model mass-luminosity relation taken from the
solar metallicity models of Baraffe et al. (1997), while the right
panels use the empirical relation tabulated in Kroupa \& Tout (1997),
combined with the relation presented in Kroupa et al. (1993) for
brighter stars.  The empirical
mass-luminosity relation produces a dip in the luminosity function
around $M_{F814W}\sim 6$ which is not apparent in the bulge data.
Despite the differences between the empirical relation and the the 
model relation, the same general conclusions can be drawn; the slope of
the mass function at low masses must be significantly shallower than at
higher masses.

We did many additional experiments to find models which match the data
to within statistical uncertainties, and we found we had to go to
models with several different power law segments to find acceptable
fits. Finding a best fit with many free parameters does not strike us
as providing significant physical insight, particularly given the
uncertainties in the binary fraction, the mass-luminosity relation, and
possible errors in our completeness correction.  We choose here to show
just several plausible mass functions which match the observed
luminosity function.  Figure 10 shows one luminosity function derived
using a mass function which has $\alpha =-2.2$ for $M>0.7$ \Msun,
$\alpha=-0.9$ for $M<0.7$\Msun, a solar metallicity model, and a binary
fraction of 0.0, and another with a binary fraction of 0.5 and mass
function slopes of $\alpha =-2.2$ for $M>0.7$ \Msun, $\alpha=-1.3$ for
$M<0.7$\Msun.

If one compares the model luminosity functions with the F555W data
(which may have less well determined completeness corrections), one
reaches similar conclusions; in fact, an even shallower faint-end slope
is required to match these data.

\section{Summary}

We have measured a deep luminosity function in the Galactic bulge, and
used it to infer a mass function. We find that the luminosity function
down to $M_I\sim 9$ is similar to that observed in the solar
neighborhood.  Transforming the luminosity function into a mass
function, we find strong evidence of a break from a power law mass
function around $0.5-0.7$\Msun.  Detailed modelling of a population
allowing for binaries and photometric errors as inferred from our data
suggests a mass function which flattens from a slope of $\alpha=-2.2$
for $M>0.7$ \Msun\  to $\alpha \sim -1$ for $M<0.7$ \Msun.  The exact
details of the derived mass function depend on assumptions about the
binary fraction, the mass-luminosity relation, and the details of our
completeness corrections.

The similarity of the mass function in the bulge to that of the solar
neighborhood is perhaps not surprising given that the mean
metallicities of the two populations may not differ by a large amount
(McWilliam \& Rich 1994).  The current data suggest that the physical
processes of star formation in the bulge and in the disk may be
similar.

Additionally, the lack of large numbers of low mass stars in the bulge
may lead to difficulties in explaining the relatively high optical
depth in microlensing events and the large number of short duration
events observed in the direction of the bulge. Models that account for
these with a population of stars require a stellar mass function with
$\alpha \sim -2$ all the way down to the hydrogen-burning limit
(Zhao, Spergel, \& Rich 1995; Han \& Gould 1996). Model luminosity functions
for this slope are shown as the steepest curves in Figure 9. If one
were to normalize these curves to the bright end of the luminosity
function, one can see that the observed star counts at the fainter
magnitudes fall significantly short of those expected for this mass 
function.

We can estimate the mass surface density observed towards Baade's
Window using the models which do a reasonable job of fitting the
observed luminosity function (Figure 10).  We compute surface mass
densities towards Baade's window for our assumed distance of 8 kpc,
using several different assumptions about the lower mass cutoff of
objects. For the model with no binaries and $\alpha=-0.9$ at the low
mass end, we derive mass densities of 1.0, 1.3, \& 1.4 $\times 10^3$
\Msun pc$^{-2}$ for lower mass cutoffs of 0.3, 0.08, and 0 \Msun.  For
the model with 50\% binaries and $\alpha=-1.3$ at the low mass end, we
derive mass densities of 1.1, 1.7, \& 2.1 $\times 10^3$ \Msun
pc$^{-2}$. If we assume that the entire bulge has a similar mass
function, we can derive a total bulge mass by scaling these numbers by
the ratio of the integrated infrared light from the bulge to that from Baade's
Window (c.f. Han 1997). For the range of mass densities above, we
derive a total bulge mass of somewhere between $7.4\times 10^9$ and
$1.5\times 10^{10}$ \Msun.

This work was supported in part by NASA under contract NAS7-918 to
JPL.  We gratefully acknowledge I. Baraffe and F. Allard for
communicating some of their stellar model and atmosphere results before
publication. We thank the referee, A. Gould, for several very useful
comments and suggestions.

\pagebreak

\pagebreak

\pagestyle{empty}

\begin{figure}
\caption{The WFPC2 F814W image of the Baade's window field.}
\end{figure}


\begin{figure}
\caption{The observed WFPC2 color-magnitude diagram. The different panels
have F555W and F814W on the ordinate. Only one in every four stars is plotted.}
\plotone{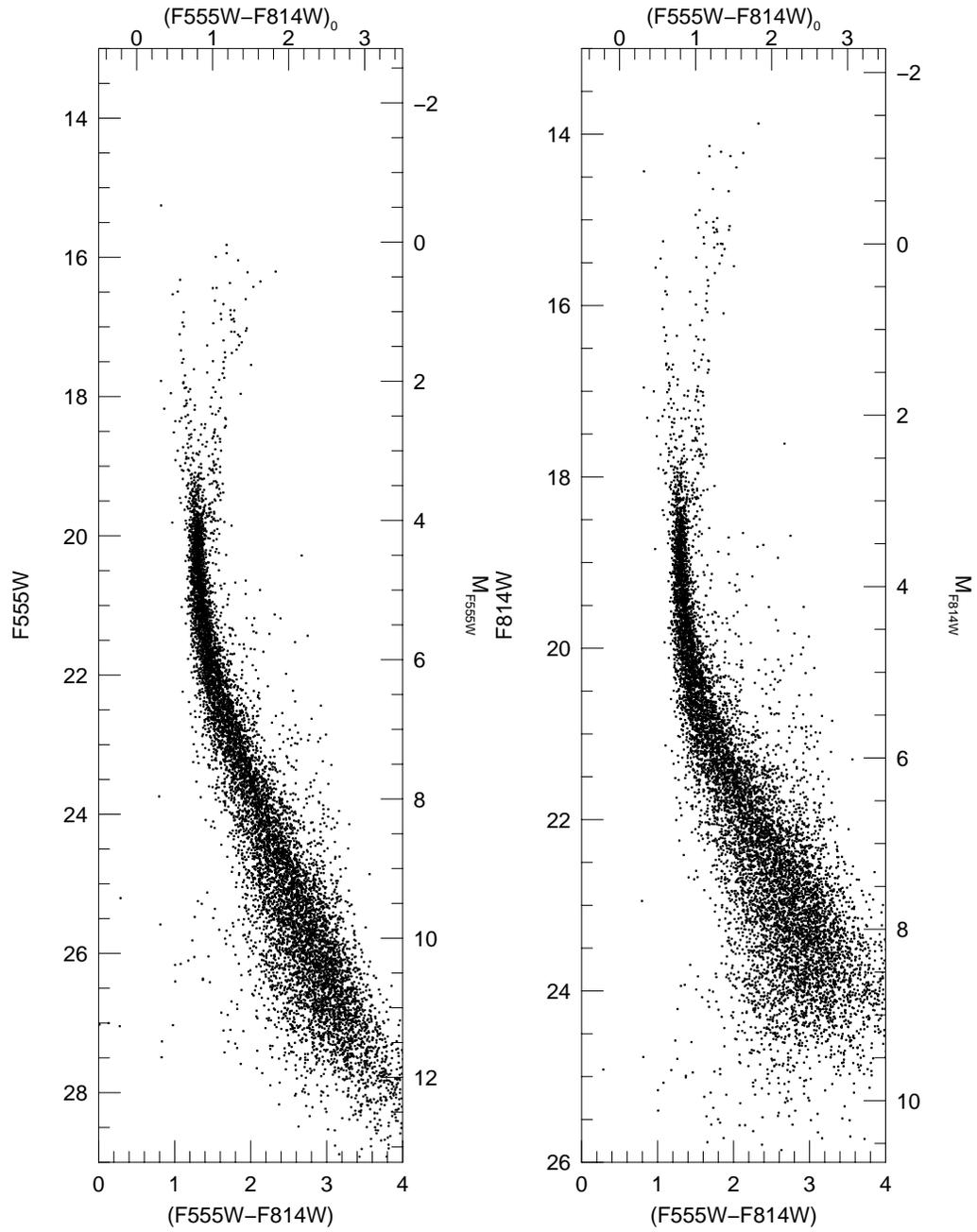}
\end{figure}

\begin{figure}
\caption{Completeness and error distributions. Each panel shows the
actual errors in F814W for simulated stars at a brightness which is given in the
upper right. The fraction of stars found at each brightness is also noted.}
\plotone{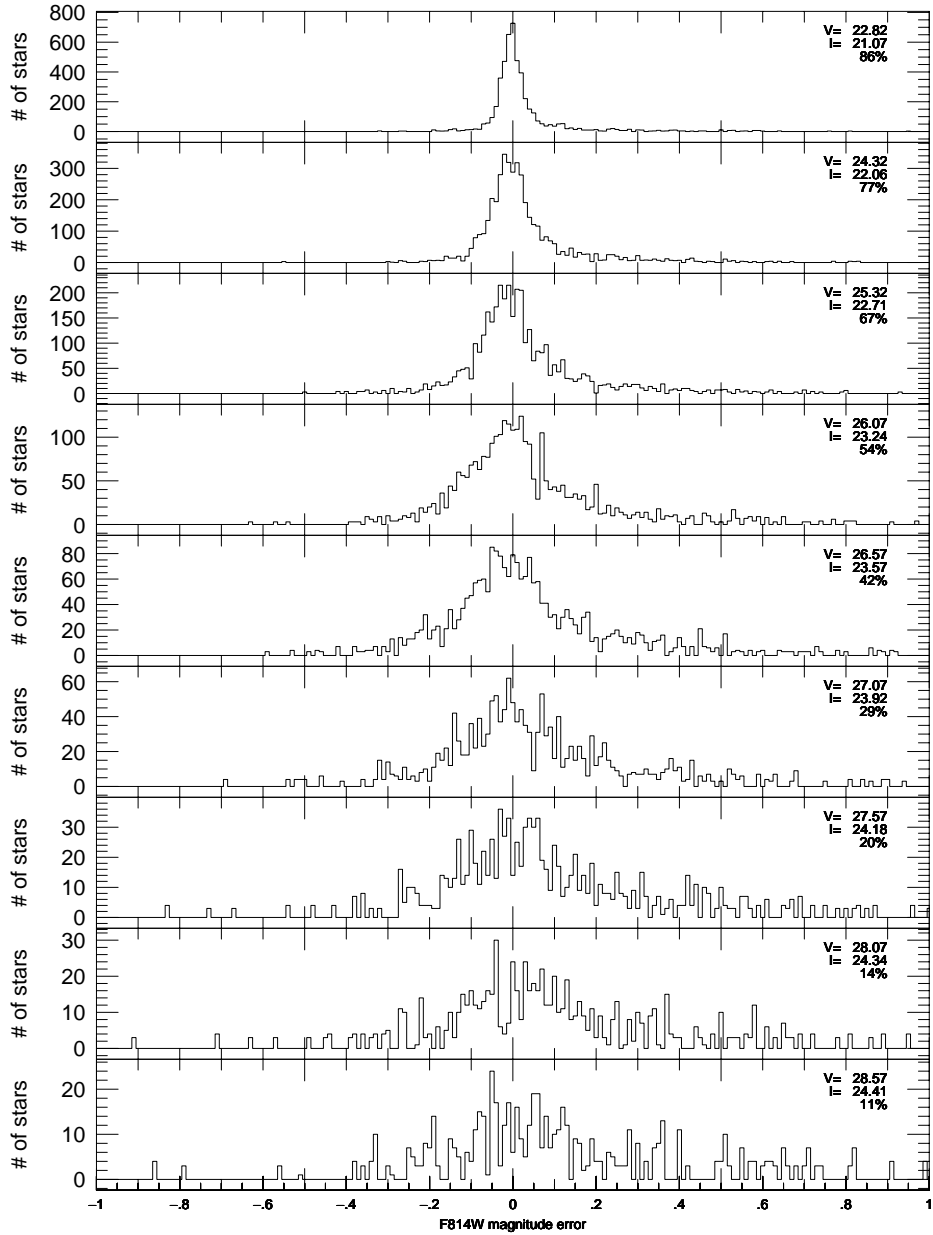}
\end{figure}

\begin{figure}
\caption{Completeness corrections, as derived from simulated stars. The
solid curve gives the correction inferred for PSFs derived to match the
actual data. The dotted curve gives the result if a different PSF is used
to create the simulated stars from that used to reduce them, where the
erroneous PSF for this case represents an extreme mismatch.}
\plotone{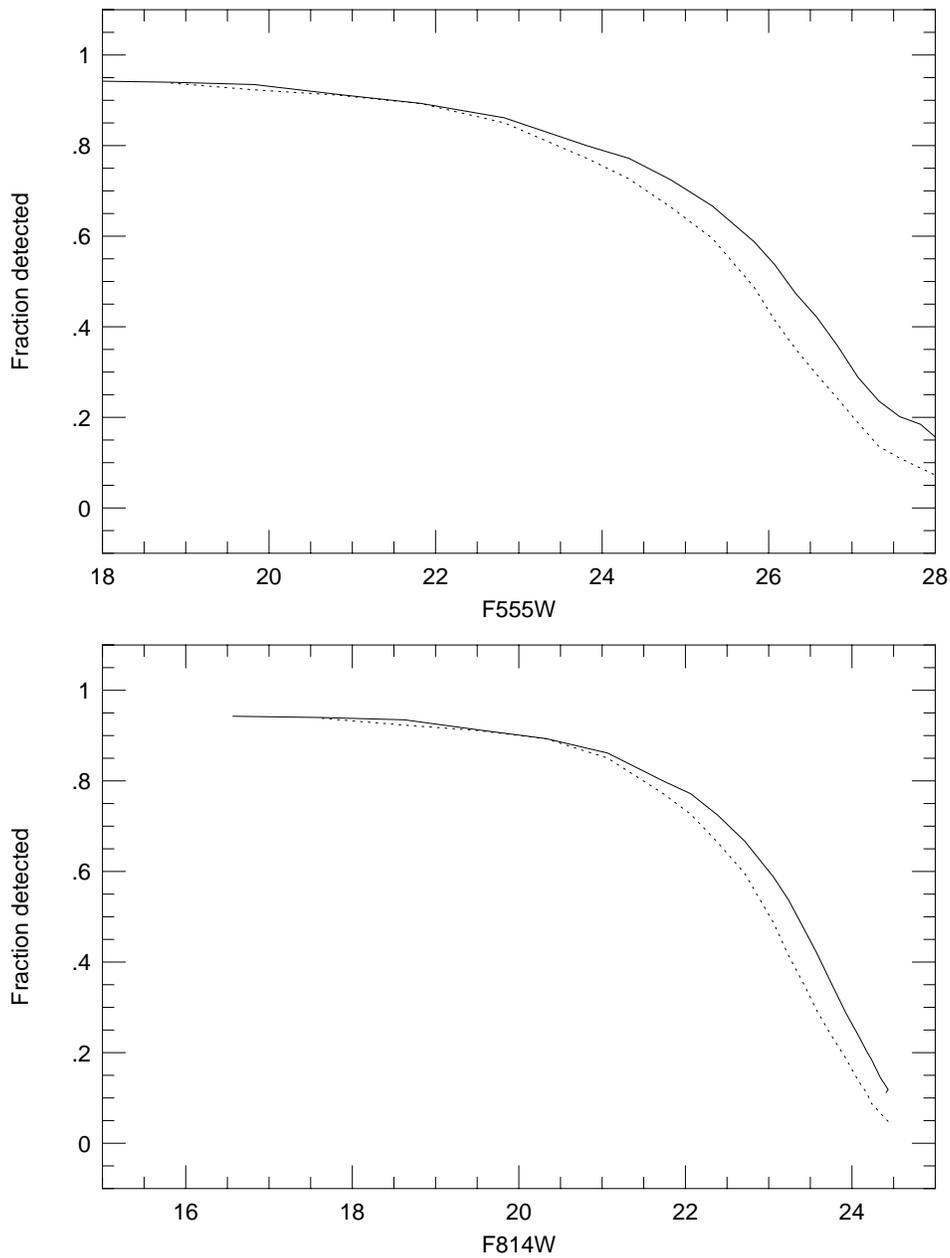}
\end{figure}

\begin{figure}
\caption{The observed WFPC2 luminosity functions, in F555W (top) and
F814W (bottom).  The raw luminosity function is shown with open
squares, while a completeness-corrected function (without correction
for systematic errors) is shown with filled squares. Data have been
normalized to number of stars per square arcmin per magnitude. Triangles show
the solar neighborhood luminosity function of Wielen et al. (1983), and
asterisks show that derived by Gould et al. (1997)}
\plotone{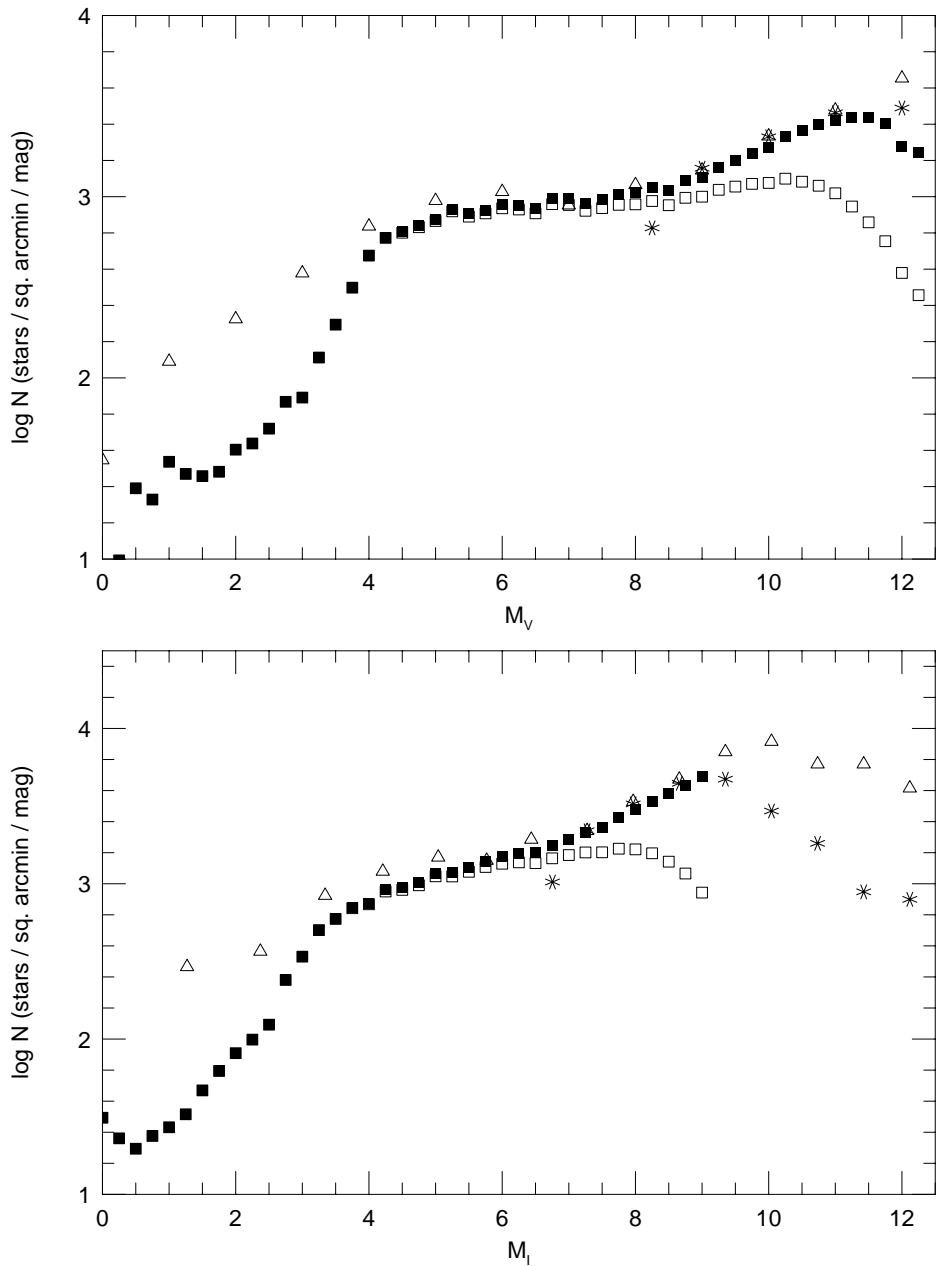}
\end{figure}

\begin{figure}
\caption{The mean locus of the bulge data in a color-magnitude diagram (points)
along with a fit to color-magnitude data of solar neighborhood stars (solid
line), and a model color-magnitude relation (dotted line). }
\plotone{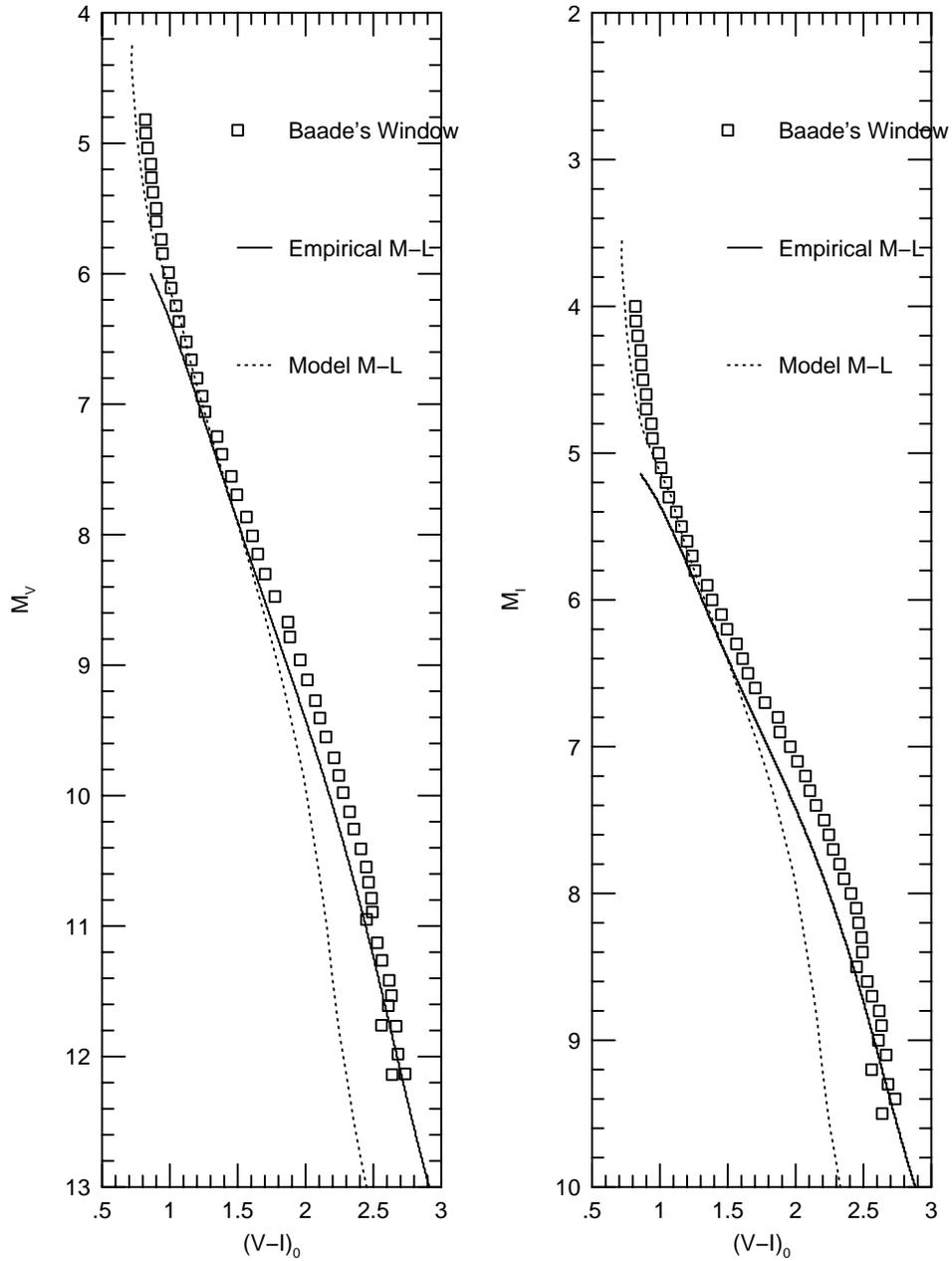}
\end{figure}

\begin{figure}
\caption{Some simulated model luminosity functions to illustrate the
effect of errors and the presence of binaries. All luminosity functions
shown here were created using the same initial mass function, with
$\alpha=-2.2$ for $M>0.5$\Msun, $\alpha=-0.5$ for $M<0.5$\Msun.}
\plotone{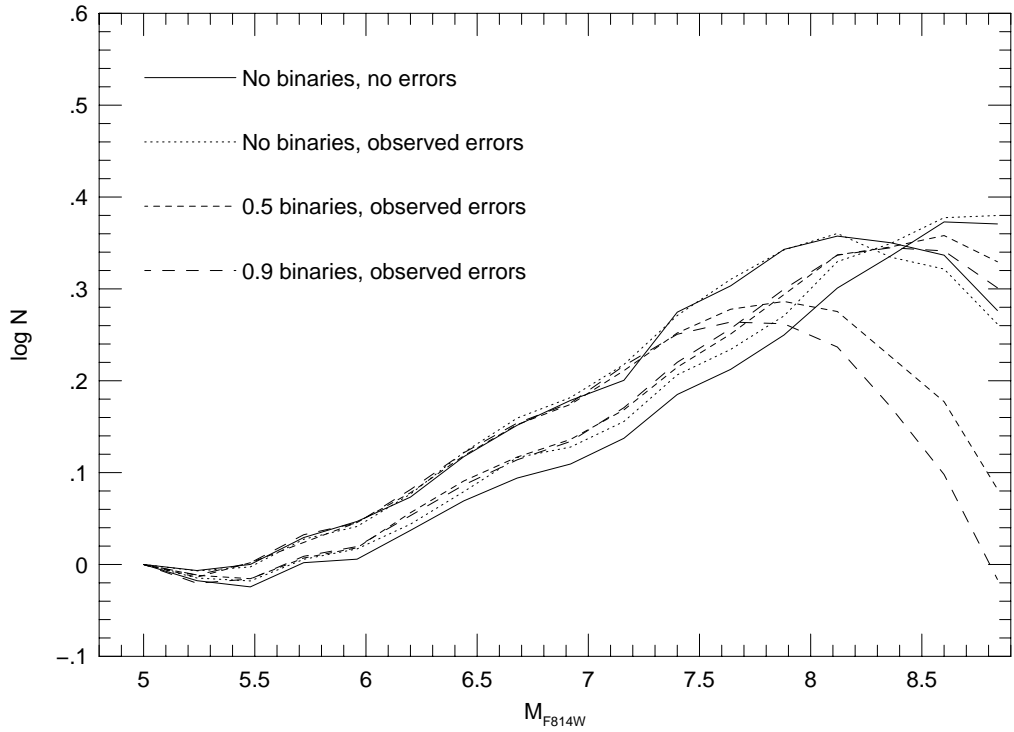}
\end{figure}

\begin{figure}
\caption{Inferred mass functions from F814W luminosity function (top)
and F555W luminosity function (bottom)
without correction for binaries or systematic errors. Results using
three different mass-luminosity relations are shown:
solar-metallicity models (triangles), lower-metalliticy 
models (circles), and empirical mass-luminosity relation (squares).
For the triangles, open symbols are for raw data, and filled
symbols show the completeness-corrected counts.
Lines above the data show slopes of $\alpha=-2.2$ and $\alpha=-0.56$.}
\plotone{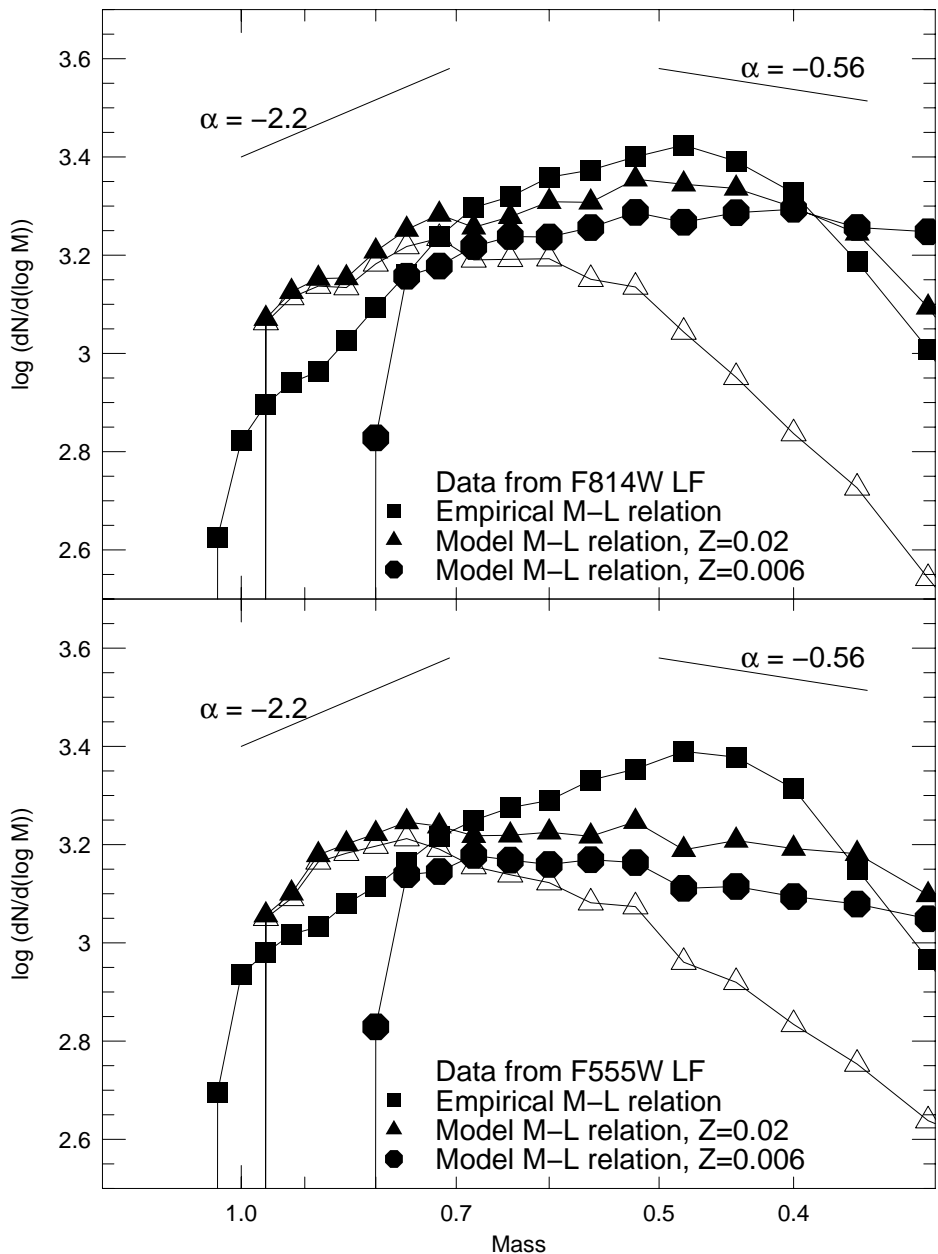}
\end{figure}

\begin{figure}
\caption{Some model F814W luminosity functions, along with the
observed one (bold line).  In each panel five models are shown. Four
models have $\alpha=-2.2$ for $M>0.5$\Msun, but different
lower-mass slopes, with $\alpha=-0.5,-0.9,-1.3,\textrm{and} -1.7$ for
$M<0.5$ \Msun. The fifth model (steepest) has a constant slope of
$\alpha=-2.0$ over the entire mass range.  Upper panels assume no
binaries, middle panels have a binary fraction of 0.5, and lower panels
have a binary fraction of 0.9.  Left panels use a solar metallicity model
mass-luminosity relation, while right panels use an empirical
mass-luminosity relation.} 
\plotone{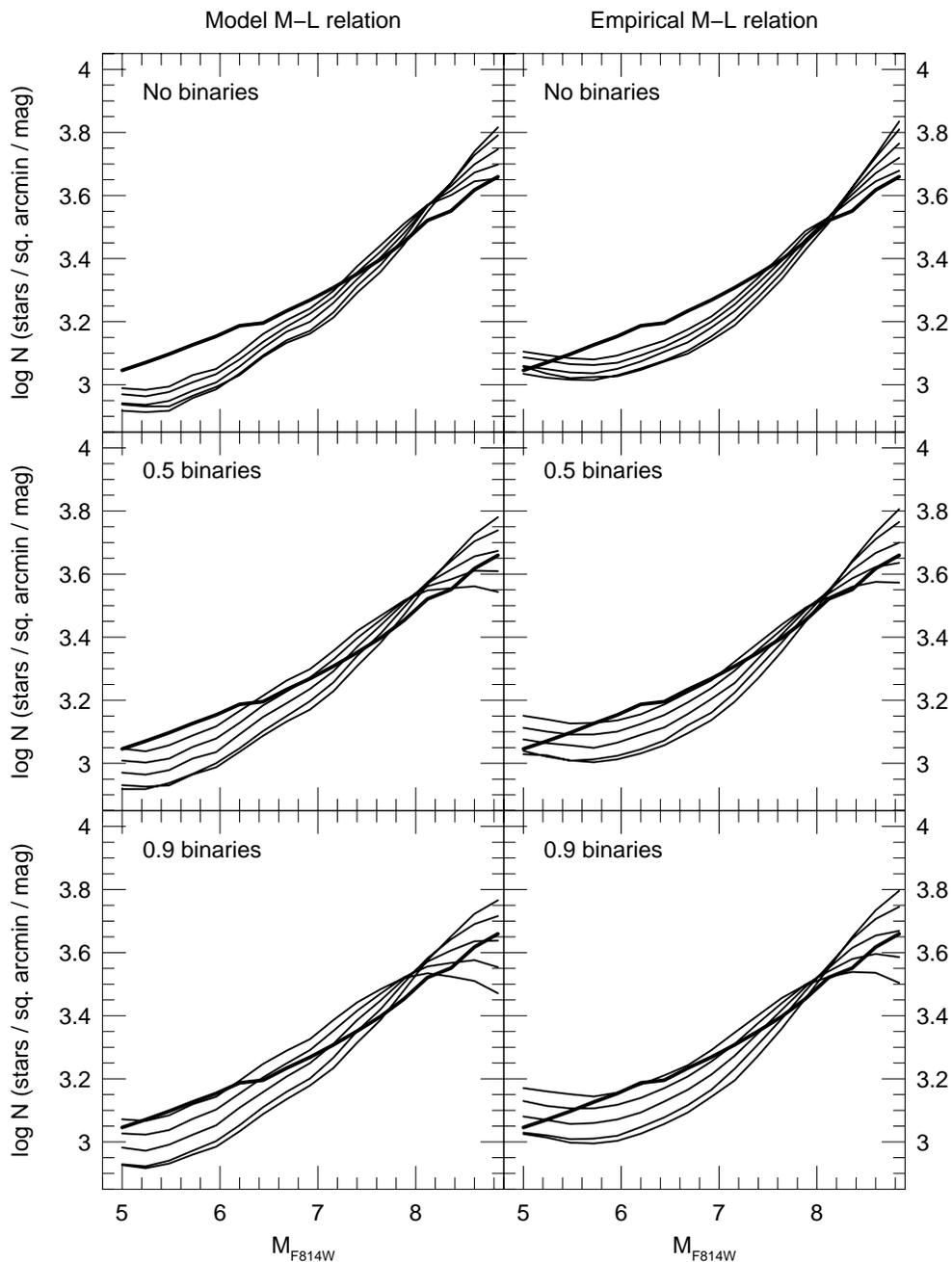} 
\end{figure}

\begin{figure}
\caption{Two reasonable-fit model luminosity functions. These use 2
different power-law segments, with a break at 0.7 \Msun.}
\plotone{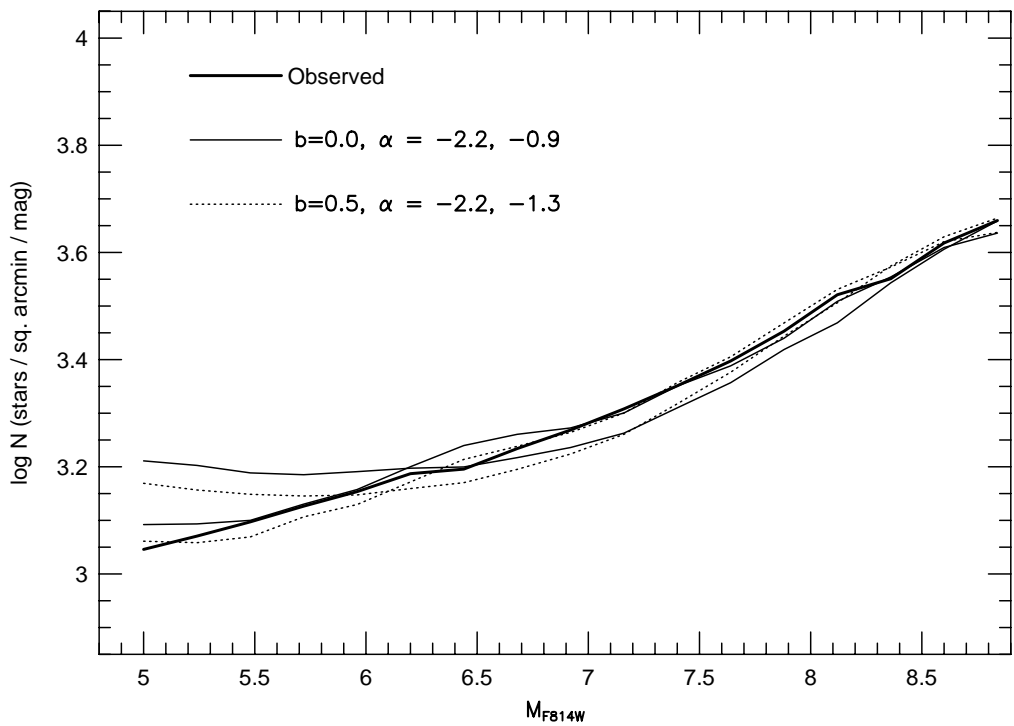}
\end{figure}


\begin{references}

\reference{} Alcock, C., Allsman, R.A., Alves, D., Axelrod, T.S., Bennett, D.P., Cook, K.H.,
Freeman, K.C., Griest, K., Guern, J., Lehner, M., Marshall, S., Park, H.-S.,
Perlmutter, S., Peterson, B.A., Pratt, M.R., Quinn, P. J., Rodgers, A. W., 
Stubbs, C. W., \& Sutherland, W., 1997, ApJ 479, 119

\reference{} Alcock, C., Allsman, R.A., Alves, D.R., Axelrod, T.S., Becker,
A.C., Bennett, D.P., Cook, K.H., Freeman, K.C., Griest, K., Gould, A., Guern,
J.A., Lehner, M.J., Marshall, S.L., Minniti, D., Peterson, B.A., Popowski, P.,
Pratt, M.R., Quinn, P.J., Rodgers, A.W., Stubbs, C.W., Sutherland, W., 
Vandehei, T., Welch, D.L., 1998, ApJ 493, in press

\reference{} Allard, F., Hauschildt, P.H., Alexander, D.R., Starrfield, S., 
1997, ARAA 35, 137

\reference{} Baraffe, I., Chabrier, G., Allarad, F., \& Hauschildt, P.H.,
1997, A\&A 327, 1054

\reference{} Bertelli, G., Bressan, A., Chiosi, C., Fagotto, F., \& Nasi, E.,
1994, AAS, 106, 275

\reference{} Bressan, A., Fagotto, F., Bertelli, G., Chiosi, C., 1993,
AAS 100, 647

\reference{} Carney, B.,  Fulbright, J. P., Terndrup, D.M., Suntzeff, N.
Walker, A., 1995, AJ 110, 1674

\reference{} Dahn, C., Lieber, J., Harris, H, \& Guetter, H.C., 1995, in An ESO
Workshop on: The Bottom of the Main Sequence and Beyond, ed. C.G. Tinney 
(Berlin, Springer)

\reference{} De Marchi, G. \& Paresce, F., 1995a, AA 304, 202

\reference{} De Marchi, G. \& Paresce, F., 1995b, AA 304, 211

\reference{} De Marchi, G. \& Paresce, F., 1997, ApJ 476, L19

\reference{} Fagotto, F., Bressan, A., Bertelli, G., \& Chiosi, C., 1994a,
AAS, 104, 365

\reference{} Fagotto, F., Bressan, A., Bertelli, G., \& Chiosi, C., 1994b,
AAS, 105, 29

\reference{} Gould, A., Bahcall, J.N., \& Flynn, C., 1997, ApJ 482, 913

\reference{} Gould, A., Popowski, P., \& Terndrup, D.M. 1998, ApJ, 492,
in press

\reference{} Gould, A., Flynn C., \& Bahcall, J.N., 1997, astro-ph 9711263

\reference{} Grillmair, C.J, Mould, J.R., Holtzman, J.A., Worthey, G.S., et al.,
ApJ, in press

\reference{} Han, C., 1997, ApJ 484, 555

\reference{} Han, C. \& Gould, A., 1996, ApJ 467, 540

\reference{} Henry, T.J. \& McCarthy, D.W, 1993, AJ, 106, 773

\reference{} Holtzman, J.A., Light, R.M., Baum, W.A., Worthey, G., Faber, 
S.M., Hunter, D.A., O'Neil, E.J., Kreidl, T.J., Groth, E.J., and Westphal, J.A.,
1993, AJ, 106, 1826

\reference{} Holtzman, J.~A., Hester, J.~J., Casertano, S., Trauger,
J.~T., Watson, A.~M., et~al.\ 1995a, PASP, 107, 156 (H95A)

\reference{} Holtzman, J.~A., Burrows, C.~J., Casertano, S., Hester, J.J.,
Trauger, J.T., Watson, A.~M., \& Worthy, G., 1995b, PASP, 107, 1065 (H95B)

\reference{} Holtzman, J.A., Mould, J.R., Gallagher, J.S., Watson, A.M \&
the WFPC2 IDT, 1997, AJ 113, 656

\reference{} Hunter, D.A., Light, R.M., Holtzman, J. A., Lynds, R., O'Neil Jr.,
E.J., \& Grillmair, C.J., 1997, AJ 478, 124

\reference{} Kroupa, P., Tout, C. A., \& Gilmore, G. 1993, MNRAS, 262, 545

\reference{} Kroupa, P.  1995, ApJ, 453, 358

\reference{} Kroupa, P. \& Tout, C. A., 1997, MNRAS, 287, 402

\reference{} McWilliam, A. \& Rich, R.M., 1994, ApJS 91, 749

\reference{} Miller, G.E. \& Scalo, J.M., 1979, ApJS, 41, 513

\reference{} Monet, D.G, Dahn, C.C., Vrba, F.J., Harris, H.C., Pier, J.R., 
Luginbuhl, C.B., \& Ables, H.D., 1992, AJ, 103, 638

\reference{} Mould, J. R., 1996, PASP 108, 35

\reference{} Richer, H.B. \& Fahlman, G.G. 1992, Nature 358, 383

\reference{} Richer, H.B., Fahlman, G.G., Buononno, R., Fusi Pecci, F., 
Searle, L., \& Thompson, I.B., 1991, ApJ 381, 147

\reference{} Salpeter, E.E., 1955, ApJ 121, 161

\reference{} Stanek, K.Z., 1996, ApJ 460, L37

\reference{} Wielen, R., Jahreiss, H., \& Kruger, R., 1983, in Nearby Stars 
and the Stellar Luminosity Function, IAU Colloquium 76, eds. A.G. Davis Phillip 
\& A.R. Upgren (Schenectady: L. Davis Press), p. 163

\reference{} Zhao, H., Spergel, D.N., \& Rich, R.M., 1995, ApJL 440, L13

\end{references}
\end{document}